# A Longitudinal Study of Non-Voice Mobile Phone Usage by Teens from an Underserved Urban Community

Ahmad Rahmati and Lin Zhong
Technical Report TR0515-09
Dept. of Electrical & Computer Engineering, Rice University

## ABSTRACT
We report a user study of over four months on the non-voice usage of mobile phones by teens from an underserved urban community in the USA where a community-wide, open-access Wi-Fi network exists. We instrumented the phones to record quantitative information regarding their usage and location in a privacy-respecting manner. We conducted focus group meetings and interviewed participants regularly for qualitative data. We present our findings on what applications our participants used and how their usage changed over time. The findings highlight the challenges to evaluating the usability of mobile systems and the value of long-term methodologies. Based on our findings, we analyze the unique values of mobile phones, as a platform technology. Our study shows that the usage is highly mobile, location-dependent, and serves multiple social purposes for the participants. Furthermore, we present concrete findings on how to perform and analyze similar user studies on mobile phones, including four contributing factors to usage evolution, and provide guidelines for their design and evaluation.

## 1. INTRODUCTION
With increasing computing power and global deployment of cellular data services, mobile phones have become a compelling platform to provide underserved communities with access to information and communication technologies (ICT). It has become critically important to evaluate the value and limitations of mobile computing beyond traditional voice applications, and to examine the interaction between community members and mobile phones in order to optimize the systems and services for the targeted communities. Existing work is limited in both scope and methodology, often focusing on a single aspect of interaction and employing short-term studies or relying solely on self reporting. In contrast, our work and others show that an accurate assessment requires a holistic long-term assessment in real-life settings.

The objective of this study is to gain a holistic assessment of the interaction through a longitudinal field trial utilizing both qualitative and privacy-respecting quantitative data collection. To this end, we converted commercial Pocket PC phones into experimental phones and distributed them to teenagers from Pecan Park, Houston, TX, an underserved community in a major urban area in the USA. The phones we distributed are Wi-Fi capable, allowing participants to access an open 802.11 network in the community for free. We developed software so that the experimental phones continuously log information related to phone usage and context in a privacy-respecting manner. To gather qualitative data, we conducted regular focus groups and interviews. The study lasted more than four months from late 2007 to early 2008. While all participants had prior experience with voice calls and some had their own mobile phones, they had little or no prior experience with advanced phones capable of providing ICT access. This enables us to observe the usage evolution of new functionalities and applications.

Findings from this long-term field study help answer the following questions:

- What applications did the participants use? (Section 5)
- How did their usage change over time and why? (Section 6)
- Characteristics (i.e. how and where) of phone usage? (Section 7)
- How did the participants share their phones with peers? (Section 8)

With these findings, we are able to analyze the unique values mobile phones provide as a platform technology for ICT access, in comparison to PCs. Furthermore, our study highlights the challenges toward the evaluation of mobile systems and services, and suggests ways to optimize their design.

It is important to note that our findings are different in nature. Some are observations and their interpretations, backed up with evidence and applied to interpretation of other observations. Others are research hypotheses that may be subjects of future investigation.

We further acknowledge that the underserved community in which the study took place, i.e. an urban one in a developed country, can be very different from those in developing countries. Therefore, it is important to note that some of our findings should be considered only relevant to populations that have a similar socioeconomic status and similar cultural background. Yet some may be generalized to other mobile users.

The remainder of this paper is organized as follows. We first discuss related work in Section 2, provide background information on our participants and their community in Section 3, and present our research instruments and methods in Section 4. We then present our findings regarding application usage in Section 5, and present the four contributing factors to usage evolution in Section 6. We present the characteristics of phone use, including its location and social use in Section 7, and present our findings regarding the sharing of phones in Section 8. We finish with discussions and conclusions in Sections 9 and 10, respectively.

## 2. RELATED WORK
Human factors in the design of mobile phones and services have been the subject of intensive research. Our work stands out in the following ways. First, existing work often address a specific aspect of mobile phone design, such as the user interface (e.g. input methods [23] and navigation [45]), availability [37], acquisition and replacement [16], or a specific service such as text messaging [12, 13], the integrated camera [30, 31], text entry [36], and mobile video [8, 29]. Social aspects of mobile phone usage have also been widely studied [2, 17, 18, 41, 43], e.g. social significance [41] and characteristics of mobile communication [17], and a specific novel application [2]. In contrast, our focus is on the holistic usage of the mobile phone and its services for ICT access, and presents guidelines for future studies.

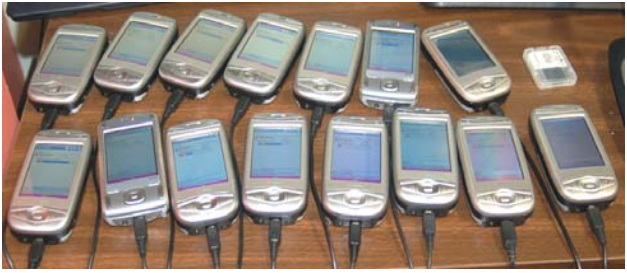

**Figure 1. Experimental mobile phones were tested in the lab before the study**

Second, existing work on human-centered mobile phone design often focuses on well-to-do communities, especially mobile professionals, e.g., [22, 28, 34]. Recently, considerable research has been devoted to mobile phones is underserved communities, often as an ICT access tool [4, 9-11, 15, 20, 21, 25, 27, 32, 33, 44]. For example, Kam *et al.* used mobile phones as a language-learning tool for children from developing countries [20, 21]. Le Dantec et al. studied the social impact of mobile phones on the homeless [25]. These works focus on either traditional voice applications, or a specific, novel application. In contrast, our work examines the generic usage of mobile phones and the values they provide beyond their traditional voice application. In addition, many studies have provided a macroscopic view of the economic impact and social significance of mobile phones [40, 42]. These studies almost unanimously acknowledge their positive impact. In contrast, our work constitutes a microscopic view into *how* teens from an underserved community interacted with mobile phones.

Third, recent work has also suggested the importance of long-term, ethnographical methods for the evaluation of mobile devices and services. For example, [3] shows that understanding the users context and culture is necessary for assessment of a mobile phone. [36] uses cameras mounted on the phone to assess device usage in naturalistic settings. These works and others [10, 14, 19, 38, 39] highlight the importance of performing user studies in real life usage, outside of lab environments. Our findings confirm the importance of long term studies in natural settings. Furthermore, we present the importance of holistic studies, and guidelines regarding how long such studies should last for usage patterns to converge.

Finally, sharing of phones, PCs, and other technology devices has also been the subject of recent research. For example, Brush et al. [5] focused on sharing home-based PCs and technical devices; and Chavan *et al.* [6] focused on conventional applications of phones: voice calls and text messaging. In contrast, we present long-term trends and findings regarding sharing non-voice applications on the phones, suggesting that PC based access control designs are inadequate for phones.

## 3. COMMUNITY AND PARTICIPANTS

Our study took place in Pecan Park, an underserved community in Houston, TX, where the average household income is below the poverty line. Approximately 13% of residents in the USA are below the poverty line, which is $10,400 for a single person family, and increases $3,600 per additional person[1]. Researchers from Rice University and Technology For All, a local non-profit organization, have installed an open-access 802.11 network covering a significant portion of the community including residential areas, public schools, and parks. See tfa.rice.edu for more information about the network.

### 3.1 Long Term Study Participants

We were able to recruit 14 teenage participants from Pecan Park for the long-term study. They were between 15 and 18 years old, either attending or had just finished high school. The participants had little or no prior experience with advanced phones. Two participants were paid hourly for assistance in scheduling focus group meetings. In the rest of this paper, we use "participants" and "primary group" to refer to them, unless otherwise indicated.

All participants had PC-based Internet access at school and a good command of Internet knowledge. They used Internet-based research for their homework, using Wikipedia and Google. They were also familiar with and use social network sites, in particular MySpace. All except one had access to PCs at home; seven had PCs devoted to them. PC ownership and regular Internet access set this underserved community, i.e. an urban one in a developed country, apart from those in developing countries.

Four of our participants had their own GSM phone plans. They simply used their SIM card in our experimental phone. For them, we provided $20 gift cards at each focus group meeting as compensation. For the other participants who didn't have their own plans, we gave them prepaid SIM cards to be used with the experimental phone and provided $25 refill cards at each focus group, equivalent to between 130 to 150 minutes. We provided several tutorial sessions to participants on how to operate the experimental phone and its various features at the beginning of the study. We also provided technical support to all participants throughout the study to ensure a smooth experience.

While we believe the phones, plans, and gift cards provided reasonable monetary, educational, and recreational incentive for participation, and we were well known among the community, it was especially difficult to recruit willing participants. While we liked, and would have benefitted from more detailed logging, it would have made it more difficult, or potentially impossible to recruit a reasonable number of participants.

### 3.2 Control Group Participants

During the course of the long-term study, new research hypotheses were formulated for which we had not collect proper data in earlier focus groups. For example, we hypothesized that a change of specific behavior and assessment was related to increased typing skill. However, we had not assessed the long-term participants' typing skill at the beginning of the study.

To compensate, we recruited an additional 10 participants from the same community who were in the same age range and had similar ICT knowledge as those in the primary group. We assume the results from the control group would be similar to what would have been attainable from the long-term group at the beginning of the study. Each control group participant participated in a single user study session, which lasted about 80 minutes and involved several participants. In each session, participants were shown the phone, and asked about their opinions about various aspects of the phones. We also tested their typing speed. We provided each participant with a $20 gift card.

### 3.3 Pilot Study Participants

To optimize the instrumentation software and formulate the initial hypotheses for the long-term study, we conducted a one-month pilot study before the long-term study. The same experimental phones were used. The key difference is that the pilot study involved 10 students from Rice University, all majoring in engineering. The pilot study further constitutes a control group for us to better understand the teen participants from the underserved community. Findings from the study were reported in [35].

**Table 1. Applications available on the experimental phones**

| | |
|---|---|
| Communication | Text Messaging (SMS), Instant Messaging (IM), Email (Outlook / Web based) |
| Recreational | Media Player, Games, Camera |
| Work / Educational | Internet Explorer (IE) |
| | Word Mobile, Excel, PowerPoint, Acrobat |
| Personal Information Management (PIM) | Address Book, Calendar, Task List |

# 4. RESEARCH INSTRUMENTS & METHODS

## 4.1 Experimental Phone

We prepared experimental phones from HTC Wizards, branded as T-Mobile MDA and Cingular 8125, and gave one to each participant in both our pilot and long-term studies (Figure 1). The Wizard is a GSM phone and allows our participants to use their own SIM card. It is Bluetooth and Wi-Fi capable and has a 2.8-inch QVGA touch-screen display. For text entry, it has a sliding hardware QWERTY keyboard in addition to a small on-screen keyboard for use with a stylus, and handwriting recognition. It can be connected to and synchronized with PCs via a USB port. We supplied the USB cable and a 1 GB MiniSD storage card with the phones. The phone takes under one minute to boot; however, the user is not required to boot the phone unless it has crashed or completely ran out of battery. Under normal usage, the phone will go in a standby mode when not used and can resume operation virtually instantaneously, at a push of a button. At the beginning of the long-term study (Summer 2007), the Wizard was one of the most feature-rich commercial Pocket PC phones. Table 1 provides a list of some of its applications and features.

## 4.2 Quantitative Data Collection

### 4.2.1 Logging Software

We developed Visual C++ based logging software that recorded the following information: 1) battery level and charging status every minute; 2) display status (on or off) every minute; and 3) signal strengths and unique MAC addresses of all visible 802.11 access points every ten minutes. Due to sensitivities with the participants and the difficulty of finding willing participants, we decided not to log the application usage. Instead, we relied on qualitative data for such information. There is also a tradeoff between logging frequency and battery lifetime. The frequencies used in the long-term study were determined based on our experience from the pilot study. The logging software reduces the standby battery lifetime of the phones from about five days to three days. The participants were informed of the battery lifetime from the beginning. The logging software runs in the background and consumes less than 200KB memory. We retrieved the data when we met participants for focus groups.

### 4.2.2 Non-Voice Application Usage

As explained in Sections 1 and 2, we are only interested in non-voice applications of mobile phones. Therefore, it is important to determine when and where they are used. Because we did not log their actual application usage due to privacy concerns, we infer the usage of non-voice applications by examining the status of the LCD screen. The experimental phones used a screen timeout interval of one minute or shorter; the screen will go off one minute after the user starts a phone call. The phone screen will stay on if the user is using a non-voice application. Therefore, we consider a series of consecutive screen-on (two minutes or more) as a non-voice session. This method cannot distinguish non-voice sessions shorter than one minute from voice applications. Therefore, our data only represent non-voice sessions over one minute. Moreover, if the user makes multiple consecutive short phone calls in a period of over a minute, our method may mistake them for a non-voice session. Despite these limitations, the estimation constitutes a consistent measure of non-voice application usage. In the rest of the paper, we use applications to refer to non-voice applications, unless otherwise indicated.

## 4.3 Qualitative Data Collection

We conducted two focus groups during the one month pilot group study. The results helped us formulate the initial hypotheses and plan the first focus groups in the long-term study. During the long-term study, we held two focus group meetings every three weeks; each participant could choose to attend either one of the two. Each focus group took about 70 minutes and took place at the conference room of a non-profit organization in the community with two research team members attending. The focus groups were semi-structured. Before each focus group, we prepared the topics and questions based on results from our previous study and the analysis of existing data, in particular recently collected data. We occasionally interviewed a participant if there were issues particular to him or her. All focus group conversations were recorded with the consent of the participants, transcribed, and used alongside our notes for manual coding. In addition to using the coded data, we often revisited the audio files for context in the later analysis of the coded data.

# 5. APPLICATIONS AND THEIR USAGE

With the quantitative and qualitative data gathered over four months, the first research question we can answer is *what non-voice applications our participants used and how their usage changed over the four months.*

## 5.1 Recreational Usage

We found that recreational applications, such as Media Player, games, and to a lesser extent the camera were the most popular type of applications on the phones. All of our participants mentioned the use of at least one recreational application in each focus group. They started using the built-in games immediately, and quickly learned to load MP3 files on the MiniSD card. By the end of the first month, most participants had music collections on the phone. Our participants reported that they primarily used these applications in their free times and often used them socially and shared them with their peers, as will be further addressed in the Social Purposes section.

Media Player remained popular throughout the study, and one participant even sold his iPod, since he "always had the phone with him". On the other hand, although most participants had found and installed new games, gaming popularity dropped towards the end of the study and they often complained that the available games were boring.

We hypothesize that the attraction of recreational applications is positively correlated with their freshness; therefore, phones should allow the users to refresh recreational applications and/or their content. Our experimental phones have very limited gaming ability and the available games can lose their freshness rather quickly. On the other hand, new phone games were costly to our participants and required advanced technical knowledge for installation. As a result, gaming lost its attraction gradually. In contrast, our participants found it easy to obtain music in standard MP3 or WMA formats and load them to the phones. Therefore, Media Player had sustained its freshness through new music content and therefore remained popular throughout our study.

**Table 2. Popular applications change during the study**

| | Beginning | Midway | Towards the end |
|---|---|---|---|
| ← | Text Messaging | | → |
| ← | IM → | | |
| | | ← Word Mobile | → |
| ← | Games | → | |
| ← | Media Player | | → |
| ← | IE | | → |
| ← Stylus text entry → | | | |
| | | ← Hardware keyboard text entry | → |

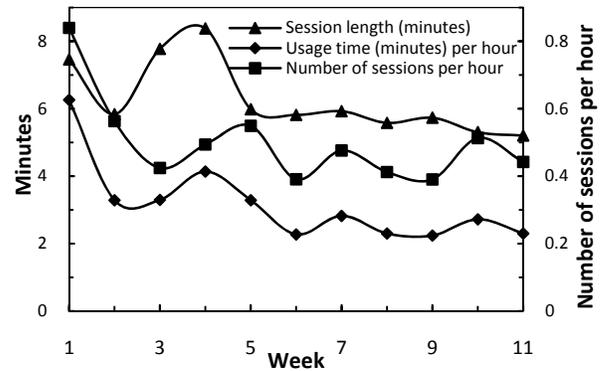

**Figure 2. Usage drops quickly, takes six weeks to stabilize**

## 5.2 Internet and Communication

### 5.2.1 Internet Connectivity
Although the experimental phones are capable of GPRS/EDGE, our participants did not have cellular data plans during the study. Instead, they had to use available Wi-Fi services for Internet connectivity, including the community open Wi-Fi network and their school Wi-Fi network. There are two technical shortcomings for Wi-Fi to provide ubiquitous wireless connectivity, in comparison with cellular data services. First, the community Wi-Fi network is intended for outdoor coverage and only one participant had usable signal inside the home. Second, Wi-Fi provides inadequate support for mobility. As a result, our participants reported disconnections when moving around outdoors. These two shortcomings presented a severe usability challenge, e.g. for Instant Messaging (IM), which will be detailed later in the Usage Evolution section.

Although our participants did not have a cellular data plan, all of them told us they would like data access and they were willing to pay for ubiquitous Internet access if the plan were cheaper; they mentioned acceptable and affordable prices as between $1 to $10 per month. At the time, cellular data plans typically cost $20 to $30 per month.

### 5.2.2 Communication Applications
Our experimental phones provided email and Instant Messaging (IM), in addition to voice communication and text messaging (SMS). During the training sessions, we showed the participants how to create an email address for those who did not have one already, and how to retrieve and send them on the phone using the included Outlook software and otherwise. However, our participants never used email for personal communication, and only occasionally used it for work related communications.

On the other hand, online social networking had become extremely popular among our target population, and all of our participants had MySpace accounts. We must note that the heavy MySpace pages were poorly supported by the phones. Our participants also reported that they regularly used IM to communicate with their friends when using a PC. Initially, they were eager to use IM on the experimental phone. But their enthusiasm disappeared a few weeks into the study due to the wireless connectivity problems mentioned above. We discuss this in further detail in the Usage Evolution section.

Our participants extensively used text messaging. Furthermore, most of them reported an increase in their amount of text messaging, indicated by changing their plan to one with an increased or unlimited number of included text messages, or by using more of their prepaid minutes for texting.

## 5.3 Work / Educational Usage
While not as popular as recreational applications, many of our participants used the phone for productivity applications and web surfing, often to fulfill their duties, such as schoolwork. By the second month of the study, they had used Word Mobile to write their homework. They used email to send their homework, and used IE to research their material. On the other hand, they did not report any use of Acrobat, Excel, or PowerPoint on the phones. According to their self reports, their usage was based on location and context, e.g., when they were in bed, or when they did not have access to PCs. For example, one participant reported that they use Word Mobile to finish late homework at school. Another one used the phone for schoolwork when he was hesitant to use a family PC due to a quarrel.

## 5.4 Usage Change
Table 2 summarizes the applications and features that were popular at different stages of the study, as described above. Note that the results were self-reported and qualitative from our focus groups. Using the logged data, we found quantitative evidence for change in usage amount. Figure 2 presents weekly statistics for non-voice phone usage by all participants in three measures: average length of usage sessions, average usage time per hour, and average number of sessions per hour. As mentioned earlier, we did not directly log our users' application usage due to privacy concerns. Instead, we infer non-voice applications based on the display status and therefore only include non-voice sessions over two minutes. Yet, Figure 2 presents strong evidence of usage change over time. First, we can see that phone usage is significantly higher in the first week, in all three measures. This suggests that the initial excitement about the phone led to increased usage. Second, we can see that it took approximately six weeks for our participants' usage to stabilize. This shows the necessity of long studies to correctly assess the usability and values of phones.

# 6. USAGE EVOLUTION

Our findings presented above clearly show that usage by our participants changed considerably over the course of the study. We have sought to find the factors influencing this evolution. While many factors can change usage, we are interested in how users explore various aspects of the phones and embrace them into their lives. Through this process, users assess the usability and usefulness of a feature and their usage changes as their assessment converges. In this section we first present four contributing factors to usage evolution and their supportive cases, and then discuss its implications in the design and evaluation of mobile devices and services.

## 6.1 Contributing Factors

Our study shows that the assessment and usage variation can take a long time to converge. Based on our observations, we have identified four qualitative, impactful contributing factors to usage evolution.

I. **Required Knowledge and Skills:** Higher requirement leads to longer assessment time because the user has to acquire the skill/knowledge to effectively use the application or feature.

II. **Context-Dependency:** Features useful or frustrating under more limited context require longer time because the user has to be in the context to experience the values and problems.

III. **Visibility and Accessibility:** More visible and accessible features require a shorter time because the user is more likely to explore them.

IV. **Initial Bias:** User assessment can be initially biased due to the users' perception about the functionality or the prestige associated with it. It takes time to remove the bias and converge on the final assessment.

We call these contributing factors to usage evolution. In the rest of this subsection, we will provide evidence supporting these contributing factors and interpret findings regarding usage evolution based on it. In the rest of the paper, we will also apply these factors when appropriate.

## 6.2 Supportive Cases

Here, we present several cases highlighting the contributing factors.

### 6.2.1 Text Entry

Text entry is an excellent example of the effect of *initial bias*, demonstrated by contrasting attitudes of participants in our long-term study and those in our control group. We had presented both stylus and hardware keyboard text entry in the training sessions. At the beginning, our long-term study participants regularly used the stylus-based text entry (on-screen keyboard and handwriting recognition) over the hardware keyboard. Similarly, our control group participants initially preferred stylus-based over hardware keyboard-based text entry after we demonstrated the three input methods. Both groups demonstrated an initial bias toward stylus-based text entry, mentioning their perception regarding its ease of use and its novelty and coolness. Further, we noticed a different bias in the long-term study participants by contrasting their attitude changes with those of our control group. All but one control group participants changed their opinion and preferred the hardware keyboard over the stylus-based text entry after they tried all three methods in the focus group, typing in a sentence. However, it took our long-term participants more than a month to really embrace the hardware keyboard, citing its accuracy. We attribute the difference to an additional bias our long-term participants may have had: the prestige gained from using stylus-based text entry before peers, who described it with comments such as "it's cool". Our control group only tried the experimental phones in the focus groups where every participant was given one to play with. They, therefore, were less likely subject to the prestige and peer opinion-based bias.

### 6.2.2 Instant Messaging (IM)

IM was a significant example of *context-dependency* in addition to *initially biased* reaction. Our participants were avid IM users on PCs. From the beginning of the study, they made serious attempts to use the IM software on the phones and aggressively sought our assistance when there was a problem in the first three weeks. Although we resolved the problem, the popularity of IM died very soon, and most participants told us they had used it "only a few times" afterwards, and they prefer to use text messaging because of limited Wi-Fi coverage and that the IM app "gets disconnected" when they move around. Indeed, over time, the novelty of IM communication with their peers had wore off and they had encountered contextual situations where IM had become frustrating to use.

### 6.2.3 Phone Charging Behavior

A clear example of the time required to *acquire knowledge* regarding certain features is the charging pattern of our participants. From our battery level and charging status logs and self-reported data from our participants, we have seen that participants learn and adapt to the battery lifetime after one to two months. For example, one of our participants regularly charged the phone during the first month. Two months into the study, he had became comfortable with the battery life and told us how the phone battery lasts a second day without charging. Our logs confirm that he charged at reduced battery levels during the third month, on average at 44%, from 63% in the first month.

### 6.2.4 Word Mobile

An important example of the time required to *acquire skills* for certain applications is our participants' assessment and use of Word Mobile. At the beginning of our study, participants reported they do not find Word Mobile useful and rarely use it. Similarly, our control group participants dismissed Word Mobile in the focus group meetings. However, towards the mid of the long-term study, our long-term participants started using it to write homework when they had no access to a PC. We believe this change in their assessment and usage of Word Mobile was related to their increased typing skill. Unfortunately, we had not quantitatively assessed their typing speed at the beginning of the study. Therefore, we used our control group as an approximate of the long-term group at the beginning of the study. We used the same procedure to measure their typing speed with the hardware keyboard. Our measurements showed that our long-term participants typed at 18 to 40 words per minute (WPM), with an average of 28, at the end of the study. In contrast, our control group typed at 14 to 24 WPM, with an average of 18, which can be a good estimate of the typing speed for the long-term study participants at the beginning of the study. The drastic difference in their typing speed explains their different attitudes toward Word Mobile, which has a high requirement in text entry.

## 6.3 Implications

The four identified contributing factors to usage evolution have multiple implications in the design and evaluation of mobile devices and services.

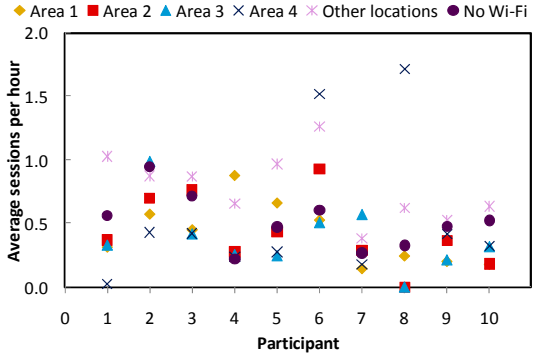
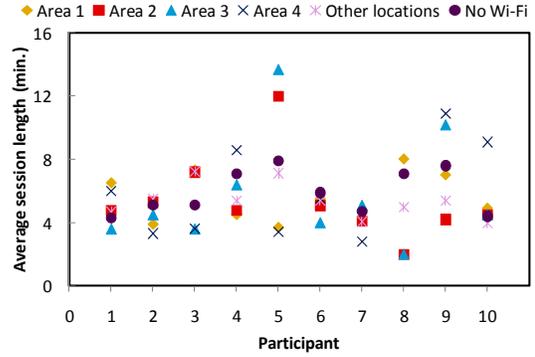

(a) Number of sessions (average = 0.42 / hour)  (b) Length of sessions (average = 6.0 minutes)

**Figure 3. Usage patterns of our participants. Areas 1 to 4 denote the top four location clusters where each participant spent their time, and were calculated for each participant separately. Locations that could not be classified due to lack of visible Wi-Fi access points are shown collectively as 'No Wi-Fi'**

*6.3.1 Evaluation*

The identified contributing factors indicate that it takes time for a mobile user to assess the usefulness and usability of a feature, which can be impacted by the required knowledge and skills, context dependency, visibility/accessibility, and initial bias of such assessment. Therefore, any evaluation must last long enough so the user gains the necessary skills and sheds their initial bias.

The identified contributing also implicate that two applications will affect each other's convergence time if they have a common skill requirement, which we call *correlated training*. Accordingly, the adoption and usage of applications and their values must not be studied independently, but in a holistic system-wide level.

Word Mobile and text messaging are an example pair of correlated applications. As noted above, our long-term participants took quite some time to develop a positive assessment of Word Mobile. At the same time, we observed their typing speed may have increased significantly, from around 18 to 28 WPM on average. While the use of Word Mobile itself could improve the typing speed, we note that text messaging, which was extensively used throughout the study, may have contributed significantly to the typing speed improvement and made our participants more comfortable with Word Mobile.

*6.3.2 Design*

The identified contributing factors indicate that it takes time for a mobile user to assess the usefulness and usability of a feature, which can be impacted by the required knowledge and skills, context dependency, visibility/accessibility, and initial bias of such assessment. Therefore, we expect that by increasing visibility, facilitating knowledge/skill acquisition, and leveraging initial bias, it is possible to promote adoption and reduce convergence time. For example, the adoption of a feature that requires an advanced skill may be facilitated by an attractive, simple game that requires the same skill but at a lower level, similar to our observation that text messaging may have improved the text entry speed and eventually promoted the adoption of Word Mobile.

## 7. CHARACTERISTICS OF PHONE USE

The small form factors and long battery lifetime make mobile phones highly accessible as they can be taken to any location and be used immediately, even on the go. Such high accessibility enables their non-voice applications to be highly mobile and used in situations where PCs may not be used. We next present findings regarding this unique feature of mobile phones and on location based usage, from both qualitative and quantitative data.

### 7.1 Physical Location Estimation

Our logging software records visible Wi-Fi access points and their signal strength every five minutes. While Wi-Fi information can be used to directly calculate approximate location [7, 24], we did not attempt to do so due to privacy considerations. Instead, to measure the location dependency of usage, we employed the Wi-Fi traces to cluster the most visited access points into areas according to their proximity. Our algorithm works as follows. Because Wi-Fi access points have a relatively short range (<~100m), we consider two access points are in the same cluster if they have been logged together many times during the entire four months (more than 40 times in our analysis) by a participant. Each cluster corresponds to a unique physical *area*, enabling us to study the location dependency of phone usage with minimal disclosure of location information.

Our method is limited to locations with visible Wi-Fi access points. Most of our participants indeed spent a significant portion of their lives in such locations; the average among all participants was 73%. To deal with locations without visible Wi-Fi access points, we cluster them together as a single area. Figure 3 (a) and (b) show, for ten participants which we collected sufficient data, the average number of non-voice sessions per hour, and the average session length at each location, respectively.

### 7.2 Mobile and Location-Dependent Usage

Due to the overhead of operating portable PCs, i.e., space requirement, startup time, and short battery lifetime, the usage of PCs is at most portable, instead of truly mobile. In contrast, we would expect significant usage of phones in all areas, even those which participants spend little time at. Our quantitative evidence shows that indeed, most participants extensively used non-voice phone applications at locations where they spent a relatively small portion of their time. These locations are aggregated and shown as 'other locations' in Figure 3. On average, our participants spent 11% of their time in all these locations, with less than 2% in each of them. Our participants had an average 0.78 sessions per hour and 5.4 minutes per session at these locations, compared to a total average of 0.42 sessions per hour and 6.0 minutes per session. Virtually all of our participants show a relatively large number of sessions per hour in these locations (Figure 3 (a)), without significantly shortening their session lengths (Figure 3 (b)). That is, they were more likely to use the phones at locations they spend less time but each time with slightly shorter duration.

While our method to infer non-voice usage is limited to usage sessions over two minutes, considering short sessions below two minutes may reveal even more unique values of mobile phones, in comparison with those of PCs, because the time overhead of turning on a PC often encourages users to choose the phone for short sessions.

Our qualitative data provide macroscopic fine-grained location information regarding usage, e.g., within homes or classrooms. Our logged data also suggest that usage was location-dependent because user objectives and usage constraints can be location dependent. We can see that the usage pattern of each user is significantly different in different areas: we observe different numbers of sessions per hour and different session lengths at different areas (Figure 3).

Our participants regularly mentioned how the high accessibility of the phone can enable truly mobile usage at a microscopic level (e.g. at different locations inside home), which is beyond the reach of portable PCs. Our qualitative data also shows that our participants leveraged the mobility of phones to facilitate Internet connectivity. While a significant part of the community is covered by the community Wi-Fi, there are many dead spots, in particular indoors. Some participants reported that they went to specific areas of their homes for better Wi-Fi signal. One of them sometimes even walked two blocks to a neighborhood park to use the free Wi-Fi network with the phone.

### 7.3 Phones Used at Home, Alongside PCs

While our location areas are calculated anonymously for each participant, we can safely assume the area where the phones spend most of their time (Area 1) is their home. Often, this includes the time when participants are asleep. We can see that for all participants, the average number of sessions per hour and their average length at home (Area 1) is comparable with their other areas. This indicates that the phones are used extensively at home, where many participants have access to a PC. This indeed corroborates with qualitative evidence from our focus groups.

Interestingly, there is no significant difference in phone usage at home between participants with devoted personal PCs and those without (Participants 3, 5, 8 and 9). This indicates that the mobile phones indeed provide unique values at home in comparison with PCs. Otherwise, better access to PCs would have led to reduced use of mobile phones.

### 7.4 Discreet Usage

The small portability, small form factor, and accessibility of mobile phones make it possible for users to access ICT in a discreet and private manner. Discreet usage refers to use under social context that conspicuous ICT access is considered inappropriate, disallowed, or simply uncomfortable.

The most prominent case is that phone usage was generally disallowed in the high schools our participants attend, except during lunch breaks. Based on the data log as reported earlier and the focus group discussion, it is obvious that this rule was routinely circumvented. We also have numerous self-reported incidents in addition to the statistics of phone usage indicating a large number of application sessions took place during school hours.

After getting the experimental phone, our participants quickly recognized and learned discreet uses. Participants in the very first focus group almost unanimously agreed that the experimental phone is difficult to hide because it is large and requires two-handed operation. By the second focus group, instead of complaints about size, we got stories indicating significant skillful discreet phone usage, which would, however, be impossible to carry out on laptops. The discreet usage includes rushing homework with the built-in Word Mobile, checking emails, text messaging, and gaming. One participant even told us how he used the phone in a way similar to a piece of paper to exchange messages back and forth within the classroom.

It is important to note that while such discreet usage may help users achieve their short-term objectives, they may be detrimental to their best interest, in particular when they are minors. It is also challenging to control discreet usage while respecting user privacy.

### 7.5 Social Motivations

The high accessibility of mobile phones allows them to be carried and used in public as well as privately, serving social purposes for their users. Previous studies have already shown mobile phones function as social symbols and fashion accessories [41]. Our experimental phones allow personalization similar to PCs in addition to ringtones and user interface options found on regular phones. Our participants leveraged this and extensively personalized the appearance and functionality of the phones; they reported use in public places and social situations so that the personalization was visible or audible. The personalization was such that from the mid of the study, we were able to tell which participant was the user of a phone just based on the phone appearance.

Our findings indicate there are two distinct types of prestige related to mobile phone usage: the first is for the possession of expensive objects and the second for the access to valued functions.

Initially, the first type of prestige, for the possession of the expensive phone, was dominant. In the first two rounds of focus groups, we got a large number of comments highlighting the perceived value of the phones among our participants' peers. For example, one participant told us "[my friends] would say 'that's a cool phone'… they really like it but it's kind of expensive."

However, we have found that prestige due to the expense of the object is short lived and quickly overshadowed by the prestige brought by having access to the functionality provided by the phones. This was indicated by responses highlighting the value of certain functionalities of the phone. For example, one participant told us "my friends still like it, because of Windows Media Player it's partly an MP3 player."

Such prestige is even more apparent when the phone carries a unique function. For example, the experimental phones have Wi-Fi capability that was uncommon even for Smartphones. One of the participants noted that the Wi-Fi capability was recognized and admired by some of her peers who own Smartphones due to the high speed of Wi-Fi compared to her peers' data plans.

### 8. SHARING

Beyond personal use as covered in Section 7, we have found that the prestige surrounding these phones induces curiosity and encourages adoption and sharing among the owners' peers.

One of the most important findings from this study is that our participants shared their experimental phones with others to achieve their social objectives. We believe mobile phones are better suited to sharing than PCs due to two reasons. First, their high accessibility brings more sharing opportunities with lower overhead, as sharing is contextual. Second, the perceived risk for phone sharing is smaller than PC sharing: sharing a car is more difficult than sharing a bicycle. This is due to fewer recognized risky usage patterns, simpler functionalities, and more straightforward restoration through resetting. Most of our participants with personal PCs told us they were wary of sharing PCs mentioning reasons ranging from fear of viruses and malware to deleting important files.

Based on our findings, we hypothesize that a user makes a decision regarding phone sharing based on perceived gain, in personal prestige and social capital, and perceived risk, in privacy and security, which are two opposite forces. The balance of the forces is dynamic, depending on the potential person to share. In the rest of this subsection, we will examine how such dynamics impact sharing behavior.

## 8.1 Evolution of Sharing

We have observed three distinct developments in the evolution of sharing among our participants

First, from the very beginning, our participants actively demonstrated the phones and their features to their peers and family. Indeed, this is a necessary step from prestige due to the possession of expensive objects to prestige due to access to valued functions, as discussed above.

Second, soon afterwards, participants started sharing the phones with their peers. Many comments from the second focus group highlighted the trend from demonstrating for friends to sharing with relatives and a few closer friends who were interested in specific features of the phone. The transition from demonstrating to sharing represents a shift from the values of prestige from conspicuous consumption to an interest in gaining social capital. While the gained prestige from the possession of the phone is short lived and represents a personal gain, the social status gained by sharing the phone with peers can be limitless and represents a gain for both the owner and the peer. Qualitative data from our focus groups indicate that our participants actively share the phones in return of social capital. We must note that the perceived social gain or loss of sharing or not sharing the device may be influenced by environmental conditions. For example, peer pressure can increase the perceived social loss of not sharing the device. In one such example, a participant told us she was worried about her personal data, but was pressured into sharing her phone with others, due to being physically small.

While our participants demonstrated the phones to practically all of their peers, they were more selective about who they shared the phones with. They unanimously mentioned privacy as the main concern that limits the circle of trusted peers they share with.

Third, as the participants used the experimental phones, they gradually "personalized" the phones through not only intentional personalization as discussed before, but also increasing amount of personal data and unintentional "traces" left in the phone, such as call history, text messages, and pictures taken with the phone camera. The privacy concern had become widespread by the second and third focus group meetings, and most participants had become more sensitive about their personal data in the phone. The personalization increases the perceived risk of sharing the phones and therefore impacts sharing. For example, one participant told us how he has to delete each text message two times (from his inbox and deleted items folders) to prevent others from accessing it when he shared the phone with them. Many participants asked for better privacy protection and access control for the wide range of private data on their phone during sharing, which is indeed unavailable on existing phones.

## 8.2 Physical Security Limits Sharing

Physical security was another factor limiting sharing of the devices. From the beginning, the participants were concerned about the physical security of the phones when sharing the phone with their peers. A few days after the start of our study, three of the phones were stolen. The stolen phones heightened the concerns of physical security. Some resorted to drastic measures. For example, one participant told us he chained the phone to his trousers for a while.

While the security concern decreased toward the middle of the study, a few isolated incidents later in the study provoked the concern again among all participants. The incidents included a SIM card being locked by friends who borrowed the phone, the headphone inadvertently broken and the phone thrown out of the window by young siblings, and the phone intentionally smashed to the ground by a classmate over quarrels.

The influence of these incidents was not just limited to the directly involved participants. The news of these incidents quickly spread at school, and our other participants also told us that they had become more wary of sharing their phones with peers and especially younger people, due to their carelessness and possibly intentional damage. Throughout the study, we observed that our participants became more selective regarding whom they could entrust the phone to, indicating a shift in balance between the personal and social values that the phone provided.

## 8.3 Design Implications

While we observed that sharing is an important way for mobile phones to provide social values to our participants, existing phones provide inadequate support for sharing with privacy assurance. When a mobile user shares their phone, they essentially give away complete access to the phone applications and data. While it may seem tempting to simply apply typical PC access control to phones, for example adding a "guest" account, it cannot support the dynamic policies that allow the owner to grant different temporary users with access to different services and data *in situ*. For example, one may want to share some photos with a family member while sharing a song with a classmate. To effectively assist sharing, the access control must support intuitive ways to specify policies, e.g., designed for one main user and multiple temporary users, and being able to quickly specify what services and data to share with a temporary user. This has motivated the design of xShare, as reported in [26].

## 9. DISCUSSIONS

The previous sections presented concrete findings from our study in the form of observation, theory, and hypothesis. We next discuss their implications in a broader context.

## 9.1 Contrast with College Students

While our study was limited in its scale for drawing statistically significant conclusions regarding underserved urban communities, there are notably differences between our teenage participants in the long-term study and college students in the pilot study.

First of all, their application usages are quite different. The most popular applications in our pilot study were Personal Information Management (PIM) applications, which were never very popular in the long-term study. On the other hand, productivity applications were hardly ever used by our pilot study participants, while they were increasingly popular to our long-term participants. There are two socioeconomic factors that likely contribute to these differences. First, their access to PCs and Internet are very different. The college participants have almost ubiquitous access to PCs and Internet when they are on campus. Therefore, the phones functioned more as a mobile extension of their PCs. In contrast, our teen participants have access to PC and Internet only at limited time and locations, even at home and school. Therefore, the Internet-capable phones assumed more importance as their ICT access, functioning similar to a PC in the pocket. Second, they have very different knowledge and prior experience with ICT access. Our college participants often had expectations biased by their PC experience. For features that are available on both phones and PCs, they often expected the phone to act similar to a PC or provide similar user experience. For example, they expected antivirus and firewall software similar to those on PCs on the phones, and some were wary of connecting to the Internet on the phones due to such security concerns. The bias introduced by such PC expectations may have hampered perceived usability for the college participants.

Another prominent difference is that our long-term study participants shared their devices with others for social purposes while none of our pilot study participants reported sharing their phones. This is apparently due to the socioeconomic status of the communities they live in. In the underserved community, Internet access and feature-rich phones are still considered privileged, there is a higher demand from the peers for sharing, and therefore sharing the device is likely to gain more personal prestige and social capital.

Moreover, our long-term participants were more concerned with the physical security of the phones than our college participants in the pilot study. While we gave every participant in both studies a belt holder for the safekeeping of the experimental phone, none of our long-term participants used them, unanimously citing the security concern for a visible phone in a belt holder. In contrast, college students in our pilot study used them and mentioned that the visible phone in the belt holder was actually fashionable. The safer community our college participants live in apparently allowed them to better benefit from the phone visibility.

## 9.2 Accessibility is Core to Unique Values

Our findings show that the experimental phones not only provide some PC services when PCs are temporarily unavailable, but also serve users in a way that is impossible for PCs. They all point to the high *accessibility* of mobile phones as the core to such services, much more accessible than portable PCs because of a smaller form factor, longer battery lifetime, and not requiring a long boot time. In particular, multiple participants reported using the phones even where they had ready access to PCs, due to the accessibility provided by not requiring a boot time.

Such high accessibility allows users to access ICT in a highly mobile fashion and use the devices in publicly visible settings, thus serving their social purposes as we discussed. In particular, the accessibility allows our participants to share their phones, often as a means for socializing and trust-building, anytime and anywhere when the context is right.

## 10. CONCLUSIONS

We present a four-month user study of mobile phone usage by 14 teens from Pecan Park, an underserved urban community in Houston, TX. We focus on the usage of non-voice applications by the participants.

We show that our participants used non-voice applications in a highly mobile fashion, at different locations and even within the same area. We observe that different usage patterns may be apply to different locations for each user, and mobile phones were used considerably when they were indeed close to PCs that were accessible to users. We show that our participants creatively leverage the high accessibility not only for ICT access but for social purposes as well. The identified contributing factors to usage evolution help us interpret our observations and presents guidelines on both designing phones and how to perform and analyze similar studies.

Our findings highlight the unique values that mobile phones, as a platform technology, provide to our participants, teenagers from an underserved urban community in the USA. Our findings strongly suggest that, at least for the studied underserved community, mobile phone design must not only address ICT needs but also serve social needs and address social concerns. We find that the long battery lifetime and the small form factor are the core to mobile phones' unique values as a platform technology for ICT access. They allow extremely high accessibility, which consequently leads to intriguing social and personal values beyond what PCs provide.

## ACKNOWLEDGEMENTS
The work was supported by NSF Award IIS/HCC #0803556. The authors would like to thank Bryan Grandy and Jim Forrest from Technology For All for their help with the user studies.